\documentclass[conference]{IEEEtran}
\pdfoutput=1
\usepackage{cite}
\usepackage{amsmath,amssymb,amsfonts}
\usepackage{graphicx}
\usepackage{textcomp}
\usepackage{listings}
\usepackage[dvipsnames]{xcolor}
\usepackage{blindtext}
\usepackage{hyperref}
\usepackage{hyperxmp}
\usepackage{caption}
\usepackage{xr}
\usepackage{microtype}
\usepackage{hyphenat}
\usepackage{soul}
\usepackage{fp}
\usepackage[most]{tcolorbox}
\usepackage{algorithm}
\usepackage{algpseudocode}
\usepackage{changepage}
\usepackage{filecontents}
\usepackage[normalem]{ulem}
\usepackage{enumitem}
\usepackage{tikz}
\usetikzlibrary{shadows, shadows.blur, arrows, automata, positioning,shapes.geometric}
\usepackage[frozencache]{minted}
\usepackage{pgf}
\usepackage{pgfplots}
\usepackage{pgfplotstable}
\usepackage{subcaption}
\usepackage{mdframed}
\usepackage{booktabs}
\usepackage{fontawesome5}
\usepackage{multirow} 

\newsavebox{\mintedbox}

\definecolor{darkgreen}{HTML}{59B02E}
\definecolor{darkred}{HTML}{B87213}

\hypersetup{
  colorlinks,
  linkcolor={ForestGreen!80!black},
  citecolor={red!70!black},
  urlcolor={blue!70!black}
}

\ifdefined\drawcomments
\newcommand{\yepang}[1]{\textcolor{blue}{[Yepang: #1]}}
\newcommand{\lili}[1]{\textcolor{cyan}{[Lili: #1]}}
\newcommand{\yuki}[1]{\textcolor{purple}{[Junfeng: #1]}}
\newcommand{\delete}[1]{\textcolor{orange}{\textst{#1}}}
\newcommand{\safedelete}[1]{\textcolor{orange}{#1}}

\newcommand{\highlight}[1]{\textcolor{magenta}{\textbf{FIXME: #1 :}}}
\newcommand{\fixed}[1]{\textcolor{darkgreen}{Fixed #1 :}}
\else
\DeclareRobustCommand{\yepang}[1]{%
\ignorespaces
}
\DeclareRobustCommand{\lili}[1]{%
\ignorespaces
}
\DeclareRobustCommand{\yuki}[1]{%
\ignorespaces
}
\DeclareRobustCommand{\delete}[1]{%
\ignorespaces
}
\DeclareRobustCommand{\safedelete}[1]{%
\ignorespaces
}
\DeclareRobustCommand{\highlight}[1]{%
\ignorespaces
}
\DeclareRobustCommand{\fixed}[1]{%
\ignorespaces
}

\fi

\definecolor{codegreen}{rgb}{0,0.6,0}
\definecolor{codegray}{rgb}{0.5,0.5,0.5}
\definecolor{codepurple}{rgb}{0.58,0,0.82}
\definecolor{backcolour}{rgb}{0.95,0.95,0.92}

\lstdefinestyle{style_java_1}{
    backgroundcolor=\color{backcolour},   
    commentstyle=\color{codegreen},
    keywordstyle=\color{magenta},
    numberstyle=\tiny\color{codegray},
    stringstyle=\color{codepurple},
    basicstyle=\ttfamily\scriptsize,
    breakatwhitespace=false,         
    breaklines=true,                 
    captionpos=b,                    
    keepspaces=true,                 
    numbers=left,                    
    numbersep=5pt,                  
    showspaces=false,                
    showstringspaces=false,
    showtabs=false,                  
    tabsize=2,
    frame=ltb,
    framerule=0pt,
    abovecaptionskip=0.2em,
    belowcaptionskip=-0.2em
}

\pgfplotsset{width=10cm,compat=1.9}

\newtcolorbox{myframe}[1][]{
  enhanced,
  arc=0pt,
  outer arc=0pt,
  colback=white,
  boxrule=0.8pt,
  #1
}

\newcommand\cfb{142}
\newcommand\coem{47}
\newcommand\cother{8}

\newcommand\fictotal{197}
\newcommand\repototal{94}
\newcommand\datesetrepos{333}

\newcounter{ctrfictotal}
\addtocounter{ctrfictotal}{\cfb}
\addtocounter{ctrfictotal}{\coem}
\addtocounter{ctrfictotal}{\cother}
\ifnum\value{ctrfictotal}=\fictotal
\else
    \PackageError{Utils}{Inconsistency!}{help text}
\fi

\newcommand\fbcamera{80}
\newcommand\fbapi{23}
\newcommand\fbui{23}
\newcommand\fbcodec{6}
\newcommand\fbbluetooth{5}
\newcommand\fbdisplay{3}
\newcommand\fbnotification{2}
\newcommand\fbundeterministic{0}

\newcommand\oemundeterministic{1}
\newcommand\oembadger{13}
\newcommand\oembiometric{1}
\newcommand\oemcamera{5}
\newcommand\oemdisplay{2}
\newcommand\oempermission{12}
\newcommand\oemsystemservice{4}
\newcommand\oemui{9}

\newcounter{fbtotal}

\addtocounter{fbtotal}{\fbcamera}
\addtocounter{fbtotal}{\fbbluetooth}
\addtocounter{fbtotal}{\fbdisplay}
\addtocounter{fbtotal}{\fbcodec}
\addtocounter{fbtotal}{\fbapi}
\addtocounter{fbtotal}{\fbui}
\addtocounter{fbtotal}{\fbundeterministic}
\addtocounter{fbtotal}{\fbnotification}

\ifnum\value{fbtotal}=\cfb
\else
    \PackageError{Utils}{The number of Functionality Break Cases does not match}{help text}
\fi

\newcounter{oemtotal}

\addtocounter{oemtotal}{\oemundeterministic}
\addtocounter{oemtotal}{\oembadger}
\addtocounter{oemtotal}{\oembiometric}
\addtocounter{oemtotal}{\oemcamera}
\addtocounter{oemtotal}{\oemdisplay}
\addtocounter{oemtotal}{\oempermission}
\addtocounter{oemtotal}{\oemsystemservice}
\addtocounter{oemtotal}{\oemui}

\ifnum\value{oemtotal}=\coem
\else
    \PackageError{Utils}{The number of OEM Features Cases does not match}{help text}
\fi

\def\calculatePercentage#1#2{%
\expandafter{\FPdiv\result{#1}{#2}\FPmul\result{\result}{100}\FPtrunc\result{\result}{0}\result\%}%
  }
\newcommand{\calSum}[2]{%
  \the\numexpr#1+#2\relax
}

\newcommand{\replace}[2]{ #2}

\def\BibTeX{{\rm B\kern-.05em{\sc i\kern-.025em b}\kern-.08em
    T\kern-.1667em\lower.7ex\hbox{E}\kern-.125emX}}

\begin{document}

\title{Demystifying Device-specific Compatibility Issues in Android Apps}

\author{%
    \IEEEauthorblockN{%
        Junfeng Chen\IEEEauthorrefmark{4}\textsuperscript{,}\IEEEauthorrefmark{1}%
        , Kevin Li\IEEEauthorrefmark{2}%
        , Yifei Chen\IEEEauthorrefmark{2}%
        , Lili Wei\IEEEauthorrefmark{2}%
        , Yepang Liu\IEEEauthorrefmark{4}\textsuperscript{,}\IEEEauthorrefmark{1}%
    }%
    \IEEEauthorblockA{%
        \IEEEauthorrefmark{4}Research Institute of Trustworthy Autonomous Systems, Southern University of Science and Technology, Shenzhen, China%
    }%
    \IEEEauthorblockA{%
        \IEEEauthorrefmark{1}Department of Computer Science and Engineering, Southern University of Science and Technology, Shenzhen, China%
    }%
    \IEEEauthorblockA{%
        \IEEEauthorrefmark{2}Department of Electrical and Computer Engineering, McGill University, Montreal, Canada%
    }%
    \IEEEauthorblockA{%
        chenjf2020@mail.sustech.edu.cn, \{kevin.li3, Yifei.Chen\}@mail.mcgill.ca%
        \\lili.wei@mcgill.ca, liuyp1@sustech.edu.cn%
    }%
}%
\maketitle%
\footnotetext[1]{This work was done when Junfeng Chen was a visiting student at McGill University. Lili Wei and Yepang Liu are the corresponding authors.}
\begin{abstract}
The Android ecosystem is profoundly fragmented due to the frequent updates of the Android system and the prevalent customizations by mobile device manufacturers.
Previous research primarily focused on identifying and repairing evolution-induced API compatibility issues, with limited consideration of \underline{d}evice-\underline{s}pecific \underline{c}ompatibility issues (DSC issues).
To fill this gap, we conduct an empirical study of \fictotal\ DSC issues collected from \repototal\ open-source repositories on GitHub. 
We introduce a new perspective for comprehending these issues by categorizing them into two principal groups, Functionality Breaks, and OEM Features, based on their manifestations and root causes.
The functionality break issues disrupt standard Android system behaviors, lead to crashes or unexpected behaviors on specific devices, and require developers to implement workarounds to preserve the original functionality.
The OEM feature issues involve the introduction of device-specific functionalities or features beyond the basic Android system. 
\replace{Our analysis of functionality break issues reveals their diverse range of effects, notably concentrated on Camera and UI.
In comparison, the OEM feature issues mainly occur in notification badges, permission management, and UI customization.}{The different nature of functionality break issues and OEM feature issues lead to unique challenges in addressing them.}
For example, our examination of issue fixing practices highlights significant differences in addressing the two categories of DSC issues. Common solutions for functionality break issues involve calling additional APIs, substituting problematic ones, or using specific parameters, while resolving OEM feature issues often relies on Android inter-component communication methods and reflection, with additional unconventional strategies.
Such observations highlight the distinctive challenges in addressing DSC issues in Android apps and will facilitate the future development of testing and analysis tools targeting these issues.
Our study demonstrates that Functionality break and OEM feature issues have different characteristics, and future research may need to investigate them separately.

\end{abstract}

\begin{IEEEkeywords}
Android, Device-Specific, Compatibility Issue
\end{IEEEkeywords}

\section{Introduction}

Android is one of the most popular operating systems for mobile devices. 
In contrast to Apple's ownership of the proprietary iOS system and its associated ecosystem \cite{apple-ios}, Google leads the Android Open Source Project (AOSP) 
\cite{aosp} and cultivates an open ecosystem. 
Device manufacturers can customize their Android devices and systems based on the AOSP.
The significant number of device models and their divergent Android systems have caused the fragmentation of the Android ecosystem. 
Customizations made by manufacturers can be buggy or incompatible with the original AOSP, which introduces \textbf{\underline{D}evice-\underline{S}pecific \underline{C}ompatibility issues} (\textbf{DSC issues} for short) \cite{paper-tse-2020,paper-taming}. 
\replace{Apps running on these devices behave abnormally compared with other devices, such as unexpected crashing and degradation of the user experience.
The apps need to be modified to adapt to these devices for better user experiences.}{Applications affected by the DSC issue will experience unexpected behavior on
specific devices, degrading user experiences.}

In prior studies, various approaches have been proposed for characterizing, identifying, and automatically fixing compatibility issues in Android apps.
Researchers have found that these approaches usually target the API-related compatibility issues induced by Android platform evolution~\cite{paper-confdroid,paper-cid,paper-dhe,paper-zhao2022-reparing,paper-liu-replicability}. 
For example, Li et al.
proposed \textsc{CiD} to model API lifecycle and detect API-related compatibility issues in apps' code~\cite{paper-cid}.
Liu et al. proposed \textit{AndroMevol} to harvest incompatible methods and fields in vendors' systems due to the evolution of the Android platform~\cite{paper-liu-tosem23-auto}. 
In comparison, little research focused on DSC issues.
DSC issues are caused by the modifications and customizations made to Android by OEMs, which have different root causes than the evolution-induced compatibility issues.
DSC issues are  generally more challenging to detect because extensive testing on many physical devices is expensive and time-consuming~\cite{paper-tse-2020,vilkomir-testing-selection}.
To address the challenge, a line of research specifically focused on testing for DSC issues.
For instance, Fazzini et al. proposed \textsc{DiffDroid} to detect inconsistent UI displays on different Android devices~\cite{paper-cpi}.
Vilkomir et al. and Khalid et al. focused on prioritizing device selection to improve the efficiency of app testing~\cite{vilkomir-testing-selection,khalid2014-Prioritizing-testing}.
However, these studies did not investigate the nature of DSC issues.
To characterize DSC issues, Wei et al. conducted the first empirical study~\cite{paper-taming,paper-tse-2020}.
Nonetheless, their main research goal was to characterize general compatibility issues in Android apps.
They analyzed DSC issues together with API-related compatibility issues.
As a result, they failed to identify the unique challenges in detecting and handling DSC issues.
Following this work, other researchers have proposed tools for API-related compatibility issues that can also handle some DSC issues~\cite{paper-taming,paper-pivot,paper-zhao2022-reparing}.
However, they can only handle simple DSC issue cases similar to API-related ones, because of the lack of in-depth understanding of DSC issues.
In this paper, we are motivated to conduct a comprehensive study on the DSC issues, analyzing their distributions and characteristics to gain a deeper understanding of these issues.

We manually inspected the code in each repository and identified \fictotal\ DSC issues from \repototal\ repositories.
Our study aims to answer the following three research questions:

\begin{itemize}[leftmargin=*,topsep=3pt]
\item \textbf{RQ1 (Issue Types)} 
\textit{What are the common types of device-specific compatibility issues in real-world Android apps?}

\textit{Motivation}.
To understand DSC issues, we first need to categorize the issues according to their causes and manifestations.
Such a categorization will guide us to investigate and compare the characteristics of different DSC issues.
\item \textbf{RQ2 (Affected Functionalities)} 
\textit{What system and app functionalities could be adversely affected by device-specific compatibility issues?}

\textit{Motivation}.
DSC issues can potentially affect a wide range of Android functionalities to various extents.
To help developers prioritize their efforts in compatibility testing, we inspect which functionalities are most likely to be affected by DSC issues.

\item \textbf{RQ3 (Issue Fixing Practices)}
\textit{How do developers fix device-specific compatibility issues? Are there different practices in fixing different types of DSC issues?}

\textit{Motivation}.
Understanding how developers resolve DSC issues can aid in automating the analysis and repairing of these issues, providing valuable insights for enhancing Android development practices and app quality.

\end{itemize}

\textbf{Key Findings.} Via analyzing the 197 real issues, we made many observations. We highlight some important ones below.

\begin{itemize}[leftmargin=*,topsep=3pt]
\item In RQ1, we observed that DSC issues can be generally classified into two main groups: functionality breaks and OEM features. Functionality breaks refer to the failures of standard functions due to manufacturer customization, while OEM features encompass additional manufacturer-specific features built on the core Android system.%
The different nature of functionality breaks and OEM features can introduce distinct challenges in detecting and handling them.
Such distinction was neglected in previous studies, motivating us to compare the affected functionalities (RQ2) and fixing practices (RQ3) of these two types of DSC issues.

\item A wide spectrum of Android functionalities can be affected by DSC issues.
While functionality break issues can affect a great variety of Android functionalities, the majority of them (73\%) are related to Camera and UI. As for OEM feature issues, the affected functionalities are more focused  on customized permissions and UI features.
Meanwhile, different manufacturers can employ varied approaches to implement similar features.
\item Fixing practices for functionality break and OEM feature issues demonstrate significant variations. Addressing functionality break issues involves calling additional APIs (36\%), substituting problematic ones (15\%), or using standard approaches with specific parameters (23\%). Addressing OEM feature issues often relies on Android inter-component communication methods (28\%) and reflection (37\%). Unconventional strategies including adapting OEM features and permissions in the AndroidManifest and addressing compatibility issues without checking the running device model are also uncovered. %
\end{itemize}

\noindent To summarize, this paper makes the following contributions:
\begin{itemize}[topsep=3pt]
    \item We conducted an in-depth study with a dedicated focus on device-specific compatibility issues in Android apps. To our best knowledge, we are the first to do so.
\item We collected a dataset of \fictotal\ device-specific compatibility issues in \repototal\ well-maintained open-source Android repositories. The dataset is publicly available to support follow-up research~\cite{our-dataset}.
    \item We derived a new perspective to categorize device-specific compatibility issues into functionality breaks and OEM features. We also studied the affected components by these issues and their repair practices, which can facilitate future research on testing and automated analysis of device-specific compatibility issues.
\end{itemize}

\section{Background}

A compatibility issue arises when code that functions correctly on one device or system does not perform as intended on another.
Android uses an integer \textit{SDK level} to distinguish major version updates and API changes.
Upgrading Android API levels involves API additions, removals, and new restrictions, and such changes bring API-related compatibility issues~\cite{paper-cid}.
Developers avoid these API-related compatibility issues by checking which API level the applications are running on and carefully avoiding calls to APIs that do not exist or are not accessible at the current API level~\cite{paper-cid}.
Android phone manufacturers can modify Android to customize their distinctive systems. 
Manufacturers also customize the hardware, using more advanced camera modules to enhance product features and competitiveness~\cite{paper-pivot}. 
Unfortunately, manufacturers' modifications and customizations often bring compatibility issues that burden app developers.
Some vendors' implementations are problematic, breaking some features like the camera. 
They will also put restrictions on some of their APIs, usually for privacy and energy consumption reasons, forcing developers to implement dedicated workarounds.
These problems usually only occur on specific device models, so we call it \textbf{device-specific compatibility issue} (DSC issue).

\begin{lstlisting}[language=Java,caption={Examples of handling compatibility issues.},label={lst-api-comp},xleftmargin=0.25cm]
// (1) API-related
if (Build.VERSION.SDK_INT >= 23) {
  return context.getColor(id); 
  // This function is introduced at API level 23
} else {
  return context.getResources().getColor(id) ;
}

// (2) Device-specific
public void initBadger() {
  // General approach
  AbstractBadger badger = new DefaultBadger();
  if (badger != null) return;
  // If it fails, try the device-specific approaches
  if (Build.MANUFACTURER.equalsIgnoreCase("Xiaomi"))
    badger = new XiaomiHomeBadger(); // Xiaomi devices
  else if (Build.MANUFACTURER.equalsIgnoreCase("OPPO"))
    badger = new OPPOHomeBader();    // OPPO devices
  else 
    badger = null;
}
\end{lstlisting}

The Android framework provides applications with the \texttt{android.os.Build} class containing the device's build information~\cite{android-build-class}.
Checking runtime information is the most common method to avoid compatibility issues according to previous research \cite{paper-taming}.
To avoid API-related compatibility issues, the application should check the \texttt{SDK\_INT} field before calling some APIs to ensure these APIs exist in the current running Android version~\cite{paper-cid}. 
For DSC issues, developers will also check \texttt{MANUFACTURER}, \texttt{MODEL}, \texttt{DEVICE}, and other fields to determine which device their apps are running on and execute specific code on specific devices. 
The examples in Listing~\ref{lst-api-comp} show how developers handle API-related and device-specific compatibility issues differently. 
We picked these examples from Li et al.'s work~\cite{paper-cid} and a famous application~\cite{github-telegram}.
The application in the first example verifies the current Android version to prevent invoking \textit{context.getColor} under SDK level 23.
The second example adopts a general approach for the badger which should work in most cases, and then tries device-specific approaches separately for Xiaomi and OPPO devices.

\section{Empirical Study}

\subsection{Dataset Collection} \label{ssec:datacollection}

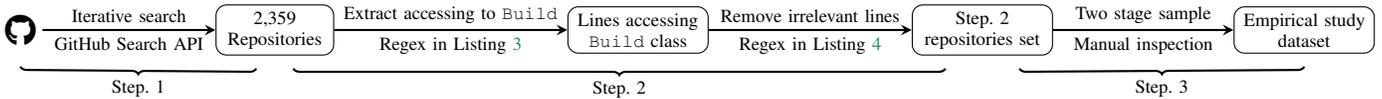
\begin{figure*}[t!h]
  \centering
  \tikzstyle{rounded-rect} = [rectangle, rounded corners, minimum width=1cm, minimum height=2em, text centered, text width=1.7cm, draw=black]
  \tikzstyle{arrow} = [thick,->,>=stealth]
  \scriptsize
  \begin{tikzpicture}[node distance=1cm]
    \node (github) [minimum height=2.5em] {\large\faIcon{github}};
    \node (repos) [rounded-rect, right=2.3cm of github, text width=1.4cm] {2,359\\Repositories};
    \node (lines) [rounded-rect, right=3.1cm of repos] {Lines accessing \texttt{Build} class};
    \node (flines) [rounded-rect, right=2.7cm of lines] {Step. 2 repositories set};
    \node (dataset) [rounded-rect, right=2.4cm of flines] {Empirical study\\dataset};

    \draw [arrow] (github) -- node[anchor=south] {Iterative search} node[anchor=north] {GitHub Search API} (repos);
    \draw [arrow] (repos) -- node[anchor=south] {Extract accessing to \texttt{Build}} node[anchor=north] {Regex in Listing~\ref{lst:regexdetect}} (lines);
    \draw [arrow] (lines) -- node[anchor=south] {Remove irrelevant lines} node[anchor=north] {Regex in Listing~\ref{lst:regexfilter}} (flines);
    \draw [arrow] (flines) -- node[anchor=south] {Two stage sample} node[anchor=north] {Manual inspection} (dataset);
    \draw [thick,decoration={brace,mirror,raise=0.7em},decorate] (github.south) -- node[anchor=north,yshift=-1em] {Step. 1} ([xshift=-1em]repos.south);
    \draw [thick,decoration={brace,mirror,raise=0.7em},decorate] ([xshift=1em]repos.south) -- node[anchor=north,yshift=-1em] {Step. 2} ([xshift=-2em]flines.south);
    \draw [thick,decoration={brace,mirror,raise=0.7em},decorate] ([xshift=2em]flines.south) -- node[anchor=north,yshift=-1em] {Step. 3} (dataset.south);
  \end{tikzpicture}
  \caption{The process of dataset collection}
  \label{fig:dataset-collection}
  \vspace{-0.5em}
\end{figure*}

\textbf{Step 1: Repository collection. }To answer the research questions, we collected open-source
repositories of Android applications or Android libraries from GitHub by searching for repositories with Android-related keywords such as ``android'' and ``android-app''.
We collected open-source repositories as our study subjects since (1) they have version tracking systems so that we can track the introduction of the code snippets handling compatibility issues, and (2) they have issue trackers where we can collect issue reports to better understand the studied compatibility issues.
To collect representative Android repositories, we also applied query filters to collect subjects that have: (1) recent updates after Jan 1, 2022 (active developing), (2) more than 50 stars (famous), and (3) more than 200 commits (well-maintained).

We leveraged the GitHub search API to collect the repositories~\cite{github-search-api}.
GitHub Search API has a limit on the number of results, returning up to 1,000 repositories per query.
We designed an iterative approach to bypass this limit and retrieve a complete repository list.
We specified in the query string to sort by the last \textit{updated} date and iteratively increased the \textit{pushed} filter, until the last returned results were less than 1,000 repositories. 
Listing~\ref{lst:searchquery} shows the template query string we used.
The \textit{keyword} we used included \textit{android}, \textit{android-app}, \textit{android-developers}, etc. %

\begin{lstlisting}[caption={The query string used in GitHub Search API},label={lst:searchquery}]
"{keyword} stars:>50 pushed:>={date} sort:updated-asc"
\end{lstlisting}

Figure~\ref{fig:dataset-collection} shows the overview of the process we used to collect the dataset and Table~\ref{tab:lines} shows the number of repositories and lines of code left in each step.
We collected these repositories in April 2023.
In the end, this step resulted in \replace{2,100}{2,359} repositories.
The script and repositories list are available in our artifacts~\cite{our-dataset}.

\textbf{Step 2: Identifying device-specific compatibility issues.} \label{par:step2} With the collected repositories, we aimed to identify the code snippets related to DSC issues. To achieve this, we aimed to extract code snippets containing accesses to device-model-related fields under the class \texttt{android.os.Build} in this research since it is an indicator of the existence of patches for DSC issues~\cite{paper-taming, paper-pivot}.
We used the regular expression in Listing~\ref{lst:regexdetect} to find lines of code accessing device information in all Java and Kotlin codes.
Although there exist unconventional methods to access device-related information~\cite{stackof-android-properties}, accessing the \texttt{andoird.os.Build} class is still the primary method adopted by most applications and libraries. We thus use the general regular expression to capture as many potential issues as possible.
In this step, we collected 3,429 lines of interest from 395 of our collected repositories.

\begin{lstlisting}[caption={The regular expression to find accesses to device-model-related fields under the class \texttt{android.os.Build}},label={lst:regexdetect}]
Build\.(BOARD|BRAND|DEVICE|MANUFACTURER|MODEL|PRODUCT)
\end{lstlisting}

We manually examined the lines of interest collected from the previous step and discovered that a number of the collected lines were not related to compatibility issues. For example, the device information is commonly used in logging and error reporting.
In these cases, the usage of the device information shares similar patterns: they are typically used in string manipulation APIs (\texttt{String.format}, \texttt{StringBuilder.append}), logging APIs (APIs in class \texttt{Log}), and concatenation with String literals.
To remove these irrelevant lines, we applied the regular expression in Listing~\ref{lst:regexfilter} to identify the previously mentioned use cases of the device-specific fields in \texttt{android.os.Build}. %
We only applied the filters to Java code because Java has a simpler syntax where it is possible to precisely filter out irrelevant cases using the regular expression without analyzing the context. %
In this step, we filtered out 842 irrelevant lines resulting in 2,587 lines from 333 repositories remaining.
Nevertheless, we could not filter out all \replace{non-FIC}{non-compatibility} issue cases using only regular expressions.
As a result, we further manually analyzed the subjects.

\begin{lstlisting}[caption={Regular expressions to identify the codes that should be filtered out},label={lst:regexfilter}]
(append|(String\.format)|(Log\.(v|d|i|w|e|f))|(\"\ *\+\ *Build\..*\+\ *\".*\")|(//.*Build.*)|(/\s*\*.*Build)|(\s*\".*\"\s*\+\s*(android\.os\.)?Build))
\end{lstlisting}

\begin{table}
  \newcommand\Tstrut{\rule{0pt}{2.6ex}}         %
  \newcommand\Bstrut{\rule[-1.2ex]{0pt}{0pt}}   %
  \caption{\replace{Lines of codes in each step}{The number of repositories and LoC in each step}}
  \label{tab:lines}
  \centering
  \setlength{\extrarowheight}{0pt}
  \begin{tabular}{clcc}
    \toprule
    Step & Dataset Processing & \# of repos & \# of LoC \\
    \midrule
    Step. 1  & Search and clone repositories & 2,359 & N/A\Bstrut\\
    \hline\Tstrut
    \multirow{2}{*}{Step. 2} & Find accessing to \texttt{Build} & 395 & 3,429\\
            & Remove irrelevant lines & 333 & 2,587\Bstrut\\
    \hline\Tstrut
    \multirow{2}{*}{Step. 3} & First stage sample & 150 & 333\\
            & Second stage sample & 285 & 1,180\\
    \bottomrule
  \end{tabular}
\vspace{-2em}
\end{table}

\textbf{Step 3: Manual inspection.} \label{par:step3} We manually inspected each line of code and its context to determine whether it was related to a DSC issue.
Specifically, we inspected whether the use of the device information caused the control flow divergence, which is a common pattern in fixes of DSC issues~\cite{paper-taming,paper-pivot}.
Three of the authors in this paper were involved in this manual analysis process.
We conducted the analysis in two phases.
In the first phase, we randomly sampled \datesetrepos\ lines of interest (this sample size to meet the statistical constraints of 95\% confidence level and 5\% confidence interval of the whole dataset).
The three authors first examined each line of interest independently.
Then, all three authors gathered and discussed the classification results to reach a consensus.
This phase helped all three authors to form a consistent understanding of the DSC issues and the categorization criteria.

In the second phase of the inspection, we extended the sample size with another 1,180 lines of interest.
We sampled 1,000 lines out of a total of 1,932 lines in Java and 180 lines out of a total of 322 lines in Kotlin.
We splited these lines into three groups.
This sample set contains 1,000 lines of Java code and 180 lines of Kotlin code.
Each author among the three authors worked in two groups independently to make sure that any sampled line was cross-checked by at least two authors.
If the two authors could not reach an agreement, the third author was involved.

By inspecting the 1,513 (333 + 1,180) lines accessing the device information in the two phases, we identified \fictotal\ device-specific compatibility issues in \repototal\ repositories.
These issues formed our empirical study dataset.

We also noted that most lines of the non-DSC issues are these cases: (1) Test code, including JUnit tests and Android Instrumentation tests. (2) Device information collection, including logging, error reporting, generating the User-Agents of HTTP requests, and uploading device information to servers.

\subsection{RQ1: Issue Types}

\subsubsection{Study Method}\label{sssec:rq1method}
In this research question, we aimed to categorize DSC issues into different types according to their root causes and manifestations.
To answer this question, we conducted a two-phase manual analysis that is similar to the procedure of Step~\hyperref[par:step3]{3} described in Section~\ref{ssec:datacollection}.
In the first phase, the three authors of the paper first independently gave a label to the type of the issue by analyzing the context information of the issues (e.g., the code, the commit messages when the lines of interest were added to the code repositories and related code comments).
The labels were then discussed to reach a consensus.
The categorization criteria of the labels were also defined in this phase.
In the second phase, the analysis was extended to the whole dataset, where each issue was inspected by at least two authors.
The third author was involved if the two authors could not reach an agreement.
In the end, if consensus could not be reached by all three authors, we assigned the issue to the ``Other'' category.
After the categorization process, the Cohen's Kappa coefficients between our three authors are 0.8872, 0.7125, and 0.8174, indicating a high level of agreement among these authors.

\subsubsection{Categorization Results}
In this RQ, we found that despite their divergent root causes and manifestations, DSC issues can be categorized into two general types, \textbf{Functionality Break} and \textbf{OEM Feature}.
Table~\ref{tab:rq1-count} shows our categorization results.
Amomg the \fictotal\ studied issues, \cfb\ (\calculatePercentage{\cfb}{\fictotal}) issues are Functionality breaks, and \coem\ (\calculatePercentage{\coem}{\fictotal}) issues are OEM features. 
We failed to agree on seven issues, and they were categorized into Other.

\begin{table}
  \caption{Identified DSC Issues}
  \centering
  \label{tab:rq1-count}
  \small
  \begin{tabular}{ccc}
     \toprule
     Type & \# of issues & \# of repositories \\
     \midrule
     Functionality break & \cfb        & 65 \\
     OEM feature         & \coem       & 36 \\
     Other               & \cother     & 7  \\
     In total            & \fictotal   & 94 \\
     \bottomrule
   \end{tabular}
\vspace{-1.7em}
\end{table}

The \textbf{Functionality Break} is a class of device-specific compatibility issues that disrupt the existing and typical approaches of the vanilla Android framework that are specified in Android Developer documentation.
Usually, this type of issue will break the existing behaviors of Android APIs.
Thus, developers have to write workarounds or use different arguments passed to APIs to avoid crashes or to keep the APIs' original behavior in affected devices.

The \textbf{OEM Feature} is another class of device-specific compatibility issues that introduce device-specific approaches to implement additional functionalities beyond its basic Android version or to utilize vendors' hardware.
To compete in the mobile phone market, manufacturers may introduce some functions specific to their own devices.
App developers adapt to these OEM features to improve user experience on specific devices.
Vendors may choose different approaches to implement the same functionality, which also introduces compatibility issues.
These issues generally require applications to leverage Android inter-component communication methods or Reflection to access the customized Android framework.

In this paper, we may use the abbreviations \textit{FB issues} and \textit{OEM issues} to refer to functionality break issues and OEM feature issues.
We selected some representative examples to demonstrate the characteristics of the two types of DSC issues.

\subsubsection{Functionality Break} \label{rq1-fb-huauwei}%

A representative example of functionality break issues is the broken FileProvider in some Huawei devices.
We identified this issue in two repositories~\cite{hw-repo-1,hw-repo-2} among our study dataset.
We also found another repository~\cite{hw-repo-3} which is not included in the study dataset but fixed the same issue differently. We classified this device-specific issue as a Functionality break because Huawei's system breaks an API contract about the storage path. Thus developers need a workaround to make FileProvider work as expected.

FileProvider is a function of the AndroidX compatibility library, which is a very commonly used library in Android development~\cite{hw-fileprovider-doc}.  
FileProvider is a special subclass of ContentProvider to share files between apps securely by creating a URI starting with \textit{content://} instead of \textit{file:///} including the absolute path.
There is an API contract described in Android documentation that the first path returned by \textit{getExternalFilesDirs} should be the same as the path returned by \textit{getExternalFilesDir}~\cite{android-doc-externalfiles}.
Huawei's system allows users to change the default external storage path to the SD card if possible.
In this case, the API \textit{getExternalFilesDir} returns the SD card path, while the API \textit{getExternalFilesDirs} returns the emulated external storage as the first path, which breaks the contract.
When FileProvider is called to share a file, it will iterate a pre-defined allow-list to check if the file is located in a permitted directory.
It takes \textit{getExternalFilesDirs[0]} as the path for external storage. If the file is located on the SD card, FileProvider will mistakenly think that the shared file is not in external storage and throw an Exception.

\begin{lstlisting}[language=Java, caption={A workaround in Stackoverflow},label=hw-provider]
public static Uri getUriForFile(Context ctx, String auth, File file) {
 if ("huawei".equalsIgnoreCase(Build.MANUFACTURER)) {
  try {
   return FileProvider.getUriForFile(ctx, auth, file);
  } catch (IllegalArgumentException e) {
   File cachePath = new File(ctx.getCacheDir(), "temp");
   File newFile = new File(cachePath, file.getName());
   FileUtils.copy(file, newFile);
   // copy the file to newFile (in the internal storage)
   return FileProvider.getUriForFile(ctx, auth, newFile);
  }
 } else return FileProvider.getUriForFile(ctx, auth, file);
}
\end{lstlisting}

A post in Stackoverflow~\cite{hw-fix-sof} proposes a simple patch (presented in Listing~\ref{hw-provider}) to fix this compatibility issue. 
It tries the default routine in all Huawei devices. 
If failed, it copies the file to the cache directory which does not cause the problem and retry.
We reproduced and confirmed this issue in a Huawei device. We extracted its Android framework and decompiled it. 
Listing~\ref{hw-provider-system-sourcecde} shows Huawei's implementation to prioritize the SD card as the default storage path by changing the returned path of API \textit{getExternalFilesDir}. 
We surmised that Huawei intends to make the application automatically adapt to SD card storage, while apps should check storage capacity and choose one by themselves according to the official Android documentation.

\begin{lstlisting}[language=Java, caption={Huawei's problematic Android Framework},label=hw-provider-system-sourcecde]
private boolean IS_SWITCH_SD_ENABLED = 
    SystemProperties.get("ro.config.switchPrimaryVolume");

private boolean checkPrimaryVolumeIsSD() {
 return SystemProperties.get("persist.sys.primarysd");
}

public File getExternalFilesDir(String type) {
 if (!IS_SWITCH_SD_ENABLED || !checkPrimaryVolumeIsSD())
   return getExternalFilesDirs(type)[0];
 if (getExternalFilesDirs(type).length == 1)
   return getExternalFilesDirs(type)[0];
 return getExternalFilesDirs(type)[1];
}
// The first path returned by getExternalFilesDirs should be the same as getExternalFilesDir
\end{lstlisting}

\subsubsection{OEM Feature}

We presented an example of an OEM introducing a similar feature before Android proposed it.
Android first proposed the function of the transparent status bar with other small flaws and proposed a repair plan after two major versions.
But other OEM manufacturers put forward their repair approaches before Android did, and removed them after Android introduced the official one, which has caused the compatibility issue that OEM-specific approaches are still required for devices running old Android.
In our emipricial study dataset, we identified 12 repositories containing codes of OEM's repair patches about this issue.

\begin{figure}[h]
   \vspace{-0.5em}
   \centering
   \captionsetup{justification=centering}
   \begin{subfigure}[b]{0.3\linewidth}
    \centering
    \includegraphics[width=\linewidth,clip,bb=0 0 1440 2560,trim=0 72cm 0 0]{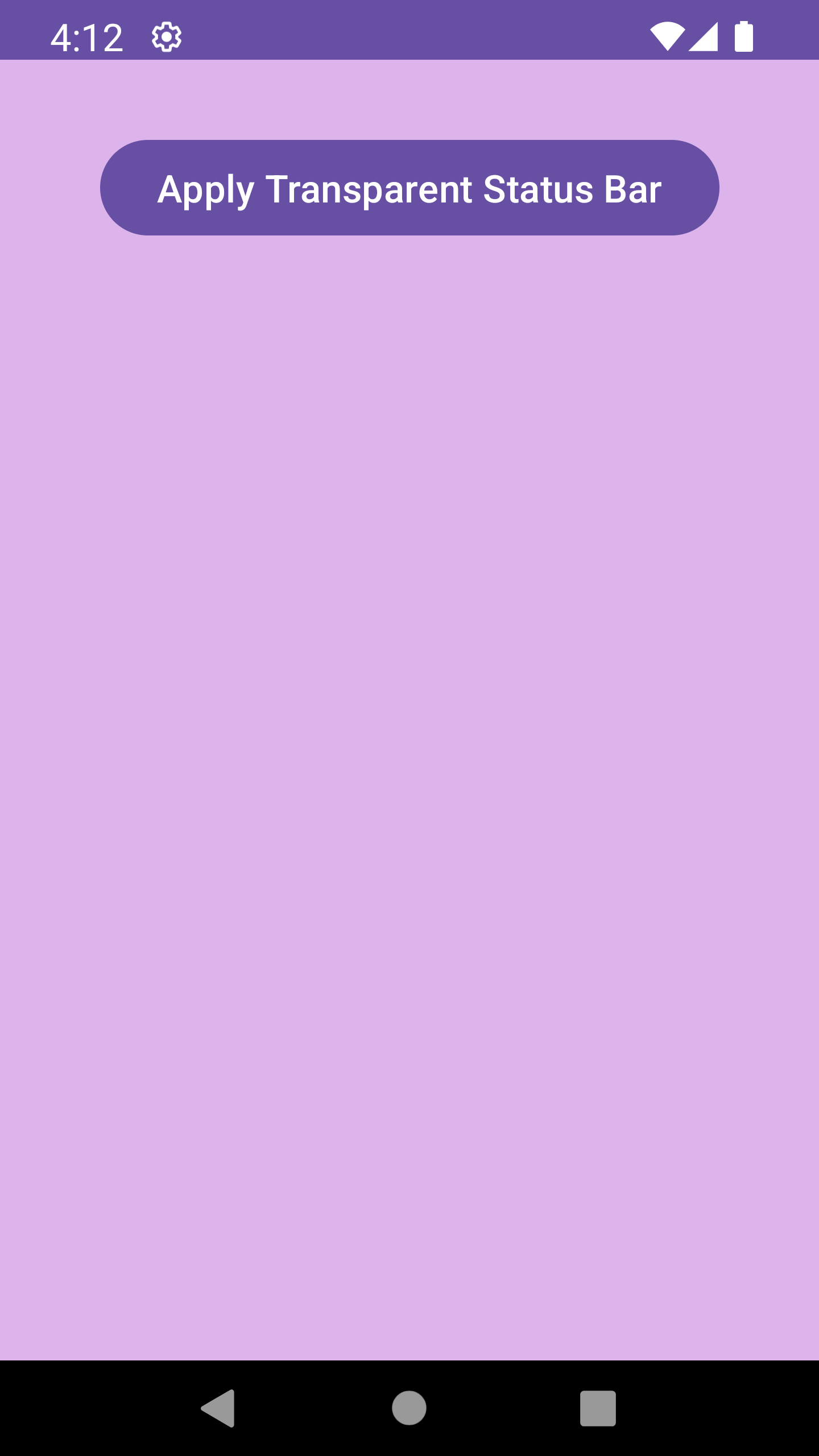}
    \caption{Original\\(API 21)}
   \end{subfigure}
   \hfill
   \begin{subfigure}[b]{0.3\linewidth}
    \centering
    \includegraphics[width=\linewidth,clip,bb=0 0 1440 2560,trim=0 72cm 0 0]{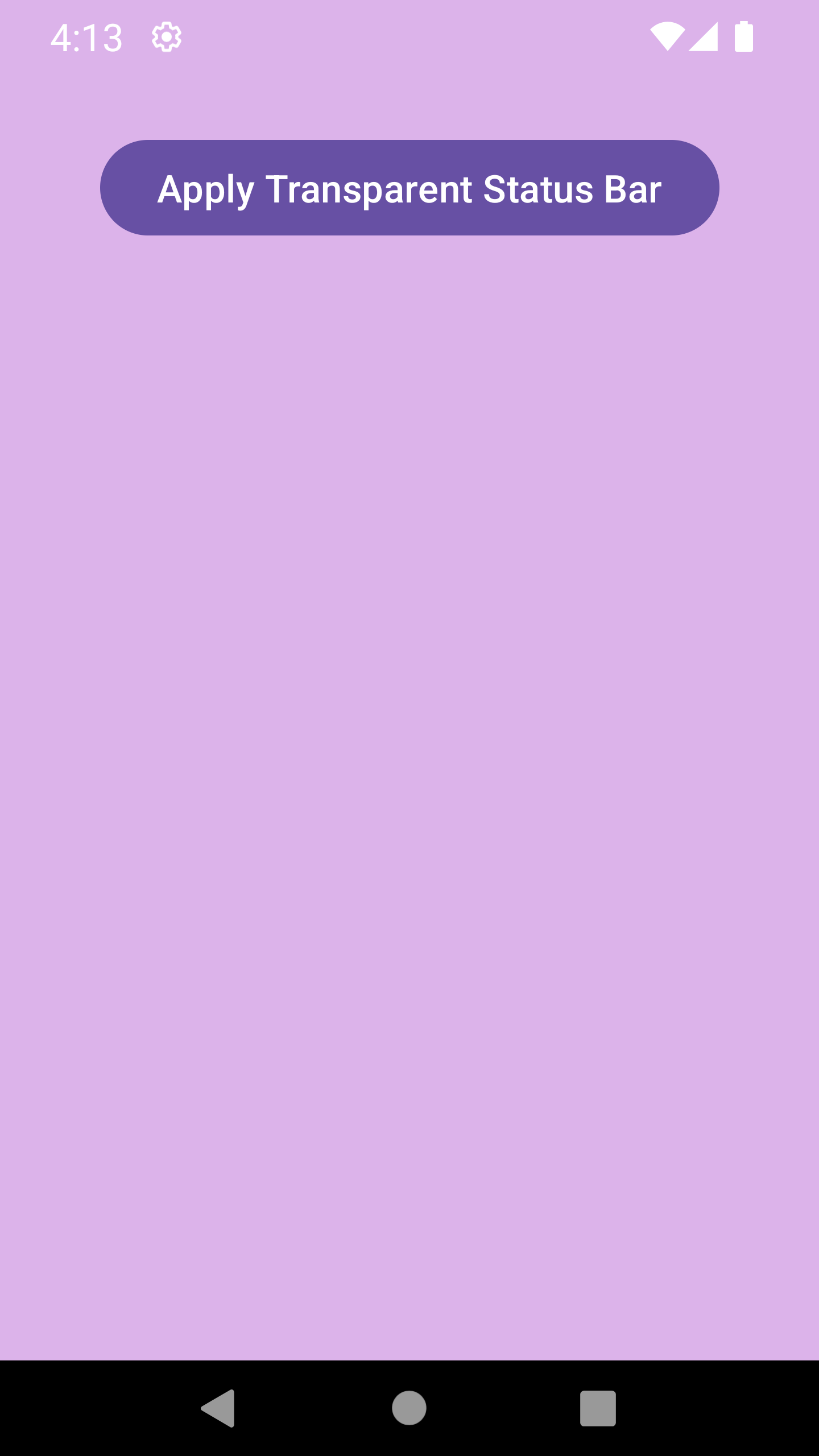}
    \caption{Transparent \\(API 21)}
   \end{subfigure}
   \hfill
   \begin{subfigure}[b]{0.3\linewidth}
    \centering
    \includegraphics[width=\linewidth,clip,bb=0 0 1440 2560,trim=0 72cm 0 0]{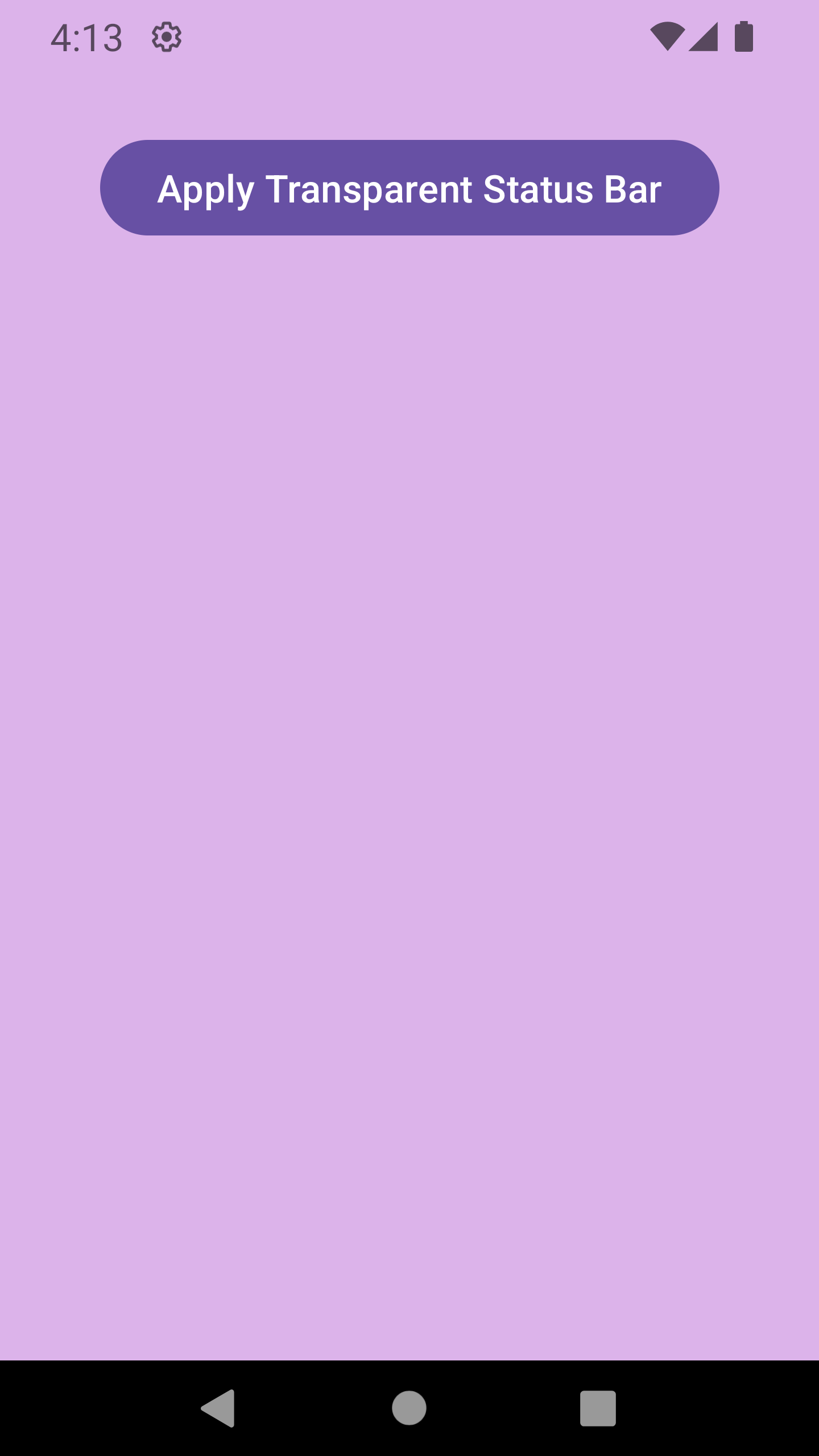}
    \caption{Light Status Bar (API 23)}
   \end{subfigure}

  %  \subfigure[\small\centering Transparent \\(API 21)]{\includegraphics[width=2.5cm,clip,bb=0 0 1440 2560,trim=0 72cm 0 0]{images/bar-21-after.png}}
  %  \quad
  %  \subfigure[\small\centering Light Status Bar (API 23)]{\includegraphics[width=2.5cm,clip,bb=0 0 1440 2560,trim=0 72cm 0 0]{images/bar-23-after.png}}
   \vspace{0.1em}
   \caption{Example of Transparent Status Bar}
   \label{fig:trans-21}
   \vspace{-0.3em}
  \end{figure}

In Android, the status bar appears at the top of the screen and contains notification icons and system icons~\cite{android-doc-statusbar}.
Android introduces a new API \textit{setStatusBarColor}~\cite{android-doc-setstatusbarcolor} in Android 5 (API 21), allowing to make the status bar transparent (Fig. \ref{fig:trans-21} (b)).
Developers may use these APIs to make the status bar transparent and display its content fullscreen for a better app experience.
This API causes another problem that applying a transparent status bar on a light background makes the white icons in the status bar challenging to distinguish.
In our example Figure \ref{fig:trans-21}(b), the background color is violet, and the icons in the upper left corner of the status bar are nearly indistinguishable.

In Android 6 (API 23), Android introduces a new View flag~\cite{android-doc-light-status-bar} to request the status bar to draw in a light mode and icons to draw in a dark way (Fig.~\ref{fig:trans-21} (c)) to fix this problem officially.
To improve user experience, OEMs have their ways to solve this problem before upgrading to API level 23.
Developers of MIUI~\cite{web-miui} are aware of this UI issue and introduced a vendor-specific approach to solving it~\cite{miui-bar-doc}.
The implementation in Listing~\ref{lst-miui} uses Reflection to access an extra field that only exists in MIUI and to call a specific method to set or clear the flag.
Listing~\ref{lst-miui} also presents another adopting method in Flyme~\cite{web-flyme} system, which is similar to MIUI's but has different parameters.
These representative cases show that apps can use Reflection to call vendor-specific interfaces, which is a common pattern in OEM issues.

\begin{lstlisting}[language=Java, caption={MIUI's and Flyme's implementation of dark status bar (Simplifed)},label=lst-miui]
void setMIUIDarkStatusBar(boolean dark, Activity act) {
 Class<?> layoutParams = Class.
    forName("android.view.MiuiWindowManager$LayoutParams");
 Field flagField = layoutParams.
    getField("EXTRA_FLAG_STATUS_BAR_DARK_MODE");
 int flag = flagField.getInt(layoutParams);
 Method setFlagMethod = act.getWindow().getClass().
    getMethod("setExtraFlags", int.class, int.class);
 setFlagMethod.invoke(act.getWindow(), dark?flag:0, flag);
}
void setFlymeDarkStatusBar(boolean dark, Activity act) {
  WindowManager.LayoutParams lp = act.getWindow().getAttributes();
  Field darkFlag = WindowManager.LayoutParams.class
    .getDeclaredField("MEIZU_FLAG_DARK_STATUS_BAR_ICON");
  int bit = darkFlag.getInt(null);
  Field meizuFlags = WindowManager.LayoutParams.class.getDeclaredField("meizuFlags");
  int value = meizuFlags.getInt(lp);
  if (dark) value |= bit; else value &= ~bit;
  meizuFlags.setInt(lp, value);
  act.getWindow().setAttributes(lp);
}
\end{lstlisting}

\begin{mdframed}[skipabove=3pt]
\textbf{Answer to RQ1:} We offered a novel perspective to understanding device-specific compatibility issues in Android apps. These issues can be broadly categorized into two main groups: functionality breaks and OEM features. The functionality break denotes the failure of standard Android functionalities due to incorrect customization by device manufacturers, whereas the OEM feature pertains to additional customized functions developed by manufacturers based on the core Android system.
The two categories of device-specific compatibility issues exhibit distinct characteristics.
\\\textbf{Implication:} FB issues and OEM issues possess different issue symptoms and root causes, showing two different areas of concern in DSC issues. 
Future research focusing on DSC issues may need to consider FB issues and OEM issues separately.
\end{mdframed}

\subsection{RQ2: Affected Functionalities}

\definecolor{col-pillar}{HTML}{5650FA}
\definecolor{col-shadow}{HTML}{816A94}
\definecolor{fillcol}{RGB}{15, 69, 115}
\pgfplotstableread[col sep=comma,]{extra-files/table.csv}\datatable
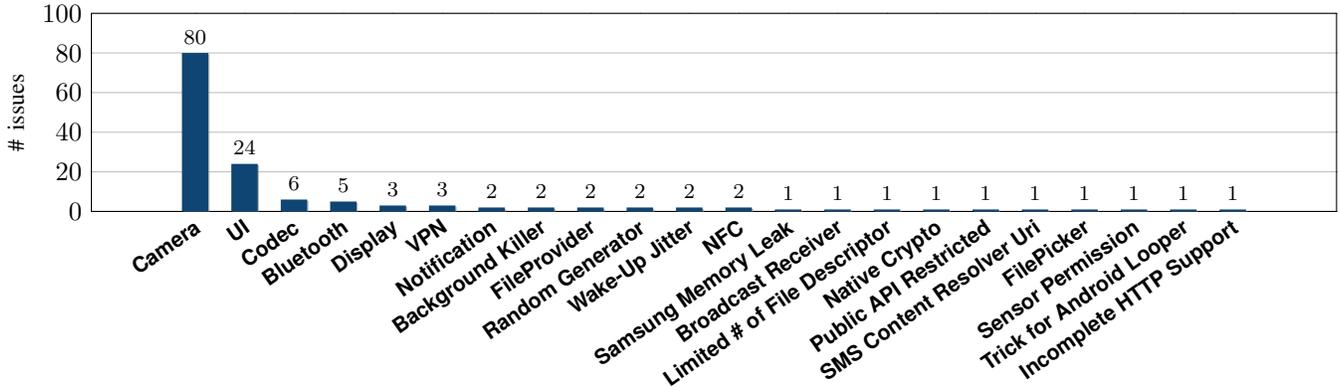
\begin{figure*}[h!]
  \centering
  \begin{tikzpicture}
    \begin{axis}[
      table/col sep=comma,
      xtick=data,
      xticklabels from table={\datatable}{Reason},
      x tick label style={font=\footnotesize\sffamily\bfseries, rotate=35, anchor=north east},
      y tick label style={font=\normalsize,major tick length=0pt},
      xticklabel shift={-2.5pt},
      width=\linewidth,
      height=12em,
      bar width=10pt,
      ymin=0,ymax=100,
      ymajorgrids,
      ylabel={\# issues},
      ylabel style={font=\small},
      clip=true,
    ]
      \addplot[
        ybar,
          visualization depends on={Count \thisrow{label}\as\mylabel},
          nodes near coords = {\mylabel{}},
          nodes near coords,
          every node near coord/.style={font=\footnotesize\sffamily,yshift=0cm,xshift=0cm},
          draw=none,
          fill=fillcol,
          drop shadow={shadow yshift=-0.5pt, shadow xshift=0.5pt},
      ] table [meta={Count}, x expr=\coordindex, y={Count}, col sep=comma] {\datatable};
    \end{axis}
  \end{tikzpicture}
  \caption{Android Functionalities Affected by Functionality Break Issues}
  \label{fig:rq2-fb}
\end{figure*}

\subsubsection{Study Method}

To better describe the characteristics of the device-specific issues, we also \replace{focus on}{inspected} the Android functionalities that are affected by each issue, such as the Camera, UI functions, Bluetooth, notification, etc. 
We determined the affected functionality of each issue based on the code context, associated Android API calls, names of their callers, and related comments.
We adopted the same cross-checking approach discussed in Section~\ref{ssec:datacollection} and Section~\ref{sssec:rq1method} to ensure that each case was checked and finally agreed upon by at least two authors. Otherwise, the case was categorized into the undetermined category.

\subsubsection{Results for Functionality Break Issues}
\texttt{ }

\replace{Table}{Figure.}~\ref{fig:rq2-fb} shows the categorization results of the affected components of functionality break issues in our study dataset.
We discovered that there is a wide spectrum of affected functionalities.
In total, we identified 22 different Android functionalities that are affected by the \cfb \ functionality break issues.
\fbcamera \ (\calculatePercentage{\fbcamera}{\cfb}) are related to Camera, and
\fbui\ (\calculatePercentage{\fbui}{\cfb}) are inconsistent UI behavior.
Other issues are scattered over 20 functionalities.
For example, \fbcodec\ (\calculatePercentage{\fbcodec}{\cfb}) are hardware MediaCodec problems, \fbbluetooth\ (\calculatePercentage{\fbbluetooth}{\cfb}) are associated with Bluetooth, and \fbdisplay\ (\calculatePercentage{\fbdisplay}{\cfb}) are associated with Android TV Display.
We now discuss examples of the major affected functionalities.

\textbf{Camera.} The camera mostly contributes to functionality-break issues (\calculatePercentage{\fbcamera}{\cfb}).
The typical pattern of camera problems is that some functions or features are inoperative. 
We delved into the specific inoperative functionalities and clustered them into the following categories based on the unusable functionality and parameters.

\begin{table}[htb]
  \caption{Affected Camera Functionalities}
  \label{tab:rq2-camera-details}
  \centering
  \footnotesize
  \begin{tabular}{lc|lc}
    \toprule
    Functionality & \# of issues  & Functionality  & \# of issues \\
    \midrule
    Capture Profile & 18 & Capture Failure & 9 \\
    Flash          & 15  & Auto Focus          & 8 \\
    Capture Parameter  & 11 & Corrupt Exif   & 7 \\
    Preview       & 10 & Corrupt JPEG Output & 2  \\
    \bottomrule
\end{tabular}
\end{table}

Table~\ref{tab:rq2-camera-details} shows our clustering result of specific inoperative camera functionalities. 
The \textit{Auto Focus} group indicates that the camera module's autofocus function is limited, e.g., continuous autofocus is disabled on some devices~\cite{web-bither-auto-focus}.
In the \textit{Flash} groups, the flash function requires additional handling, and sometimes it is necessary to use a flashlight to simulate the flash~\cite{web-androidx-isTorchAsFlash}.
In the \textit{Capture Parameter} group, some camera parameters require additional adjustments, including HDR, ZSL (Zero Shutter Lag), ISO, WB (White Balance), and AE (Auto Exposure)~\cite{web-androidx-ZslDisablerQuirk}.
In the \textit{Capture Profile} group, some photo and video quality options, such as resolution and frame rate, are not available, which can cause the encoder to crash~\cite{web-androidx-VideoEncoderCrashQuirk} or produce stretched or distorted photos or videos~\cite{web-androidx-ExcludeStretchedVideoQualityQuirk}.
In the \textit{Preview} group, the preview of a video or photo is stretched or delayed~\cite{web-androidx-PreviewDelayWhenVideoCaptureIsBoundQuirk}.
In the \textit{Capture Failure} group, some models may crash after shooting~\cite{web-androidx-CrashWhenOnDisableTooSoon}, or have no output of the image~\cite{web-androidx-ImageCaptureFailedWhenVideoCaptureIsBoundQuirk}.
In the remaining two groups, the JPEG images returned from the camera may be corrupted~\cite{androidx-camera-bug}, or the embedded Exif information is incorrect~\cite{androidx-camera-bug3}.

\textbf{UI. }UI issues are pervasive among both types of compatibility issues, especially evolution-induced. 
We identified two cases about the menu button in LG devices~\cite{web-deltachat-BaseActivity}, which have been mentioned in previous research~\cite{paper-pivot}.
This issue causes a \textit{NullPointerException} and crashes the application when the hardware menu button is pressed in LG devices.
In addition to this, we also found two issues that crash the application.
We identified a classic issue that Samsung's DatePicker implementation was broken in some of Samsung's Android 5.0 systems~\cite{samsung-datepicker-case}.
The developer used an old DatePicker before Android 5.0 to mitigate this issue.
In another issue, using \textit{TextInputEditText} from Material Components on Meizu's devices crashed the application, and further investigation revealed that the cause of the issue was a Meizu customized method that didn't take into account the possibility that an object could be null, resulting in a \textit{NullPointerException}~\cite{meizu-view-issue}. Material Components avoids this exception by calling \textit{setHint("")} to force the object non-null~\cite{meizu-fix}.

\begin{lstlisting}[language=Java,caption={ExoPlayer issue \#3249},captionpos=b,label=lst-exoplayer]
boolean isCodecUsable(MediaCodecInfo info, String codec) {
  if (SDK_INT < 24 && "OMX.Exynos.AAC.Decoder".equals(codec) && "samsung".equals(MANUFACTURER) && (DEVICE.startsWith("zeroflte") || ... )) { 
    return false;     // Galaxy S6 and other devices
  }
  // more combinations ... 
  return true;
}
\end{lstlisting}
  
\textbf{Codec}. There are also compatibility issues with media encoder/decoder or hardware-assisted functions related to multi-media.
We categorized this class of issues as Codec issues.
ExoPlayer is an application-level media player developed by Google and becomes a part of AndroidX media~\cite{exoplayer}.
Many issues from it report that some hardware encoders/decoders malfunction in particular devices.
This issue reports the Samsung Galaxy S6 series failed to decode 7.1 channel AAC audio~\cite{exoplayer-case}. Listing~\ref{lst-exoplayer} presents ExoPlayer's fix that blocks particular encoder/decoder in particular devices.
The Android client of Signal's issue page shows several devices with AEC issues~\cite{signal-webrtc}. They used to have a set of affected models hard-coded in the source code~\cite{web-signal-ApplicationContext}, but subsequent updates have enabled this list to be synchronized with the cloud.

\textbf{Others.} 
Due to the wide distribution of affected functionalities, we were unable to categorize all the FB issues.
We observed applications using the VPN Service function suffer from the limitation of Samsung's system~\cite{android-vpn-service}.
The Moviesy~\cite{ss-vpn-2} includes a patch from an Android OpenVPN client example~\cite{ss-openvpn}.
The patch states that Samsung Android 5.0+ devices will ignore DNS servers outside the VPN range, so an additional route is required for the DNS servers.

Another significant FB issue is that OEM systems have more aggressive background limitations than \replace{unmodified}{vanilla} Android, which restricts applications from running in the background, being waked by the system or other applications.
Developers from xDrip-plus complain about this issue and write a workaround to collect data from wear devices periodically~\cite{xdrip-plus}.
Generally, applications will guide users to turn off battery optimization on them and exempt them from the limitations. 
A website called \textit{Don't Kill my app!} highlights these kinds of issues and gives device-specific solutions~\cite{dont-kill-my-app}.

In other groups, Bluetooth problems are that some systems or hardware require particular Bluetooth scanning parameters. The Display issues come from some missing functionalities on Android TV devices, such as multimedia tunneling~\cite{android-multimedia-tunneling}.

\textbf{Summary of the Functionality Break category:}
It is a notable characteristic that a wide range of functions are affected by functionality break issues, including hardware-related features.
The component most susceptible to FB issues is the Camera.
Manufacturer customizations can sometimes lead to unexpected behaviors, including functionality failures or degradation, even application crashes.

\subsubsection{Results for OEM Feature Issues} ~\\
\begin{table}
  \caption{Related Android Component for OEM Feature Issues}
  \label{tab:rq2-categories-oem}
  \centering
  \small
  \begin{tabular}{ccc}
    \toprule
    Category & \# of issues & \# of repositories \\
    \midrule
    Badger      & \oembadger & 12 \\
    Permission  & \oempermission & 8 \\
    UI          & \oemui & 6 \\
    Camera      & \oemcamera & 5 \\
    System Service & \oemsystemservice & 4 \\
    Display     & \oemdisplay & 2 \\
    Biometric   & \oembiometric & 1 \\ 
    Undetermined & \oemundeterministic & 1 \\
    \bottomrule
\end{tabular}
\end{table}
Table~\ref{tab:rq2-categories-oem} shows our categorization results for OEM features.
As shown in Figure~\ref{fig:rq2-fb} and Table~\ref{tab:rq2-categories-oem}, different components are affected by functionality breaks and OEM features. In \coem\ OEM feature issues, we categorized that \oembadger\ (\calculatePercentage{\oembadger}{\coem}) of them are related to the Badger function, \oempermission\ (\calculatePercentage{\oempermission}{\coem}) are related to Permission, \oemsystemservice\ (\calculatePercentage{\oemsystemservice}{\coem}) are differences on System Service, \oemui\ (\calculatePercentage{\oemui}{\coem}) are related to UI, \oemcamera\ (\calculatePercentage{\oemcamera}{\coem}) are associated with Camera, \oemdisplay\ (\calculatePercentage{\oemdisplay}{\coem}) are associated with Display, and \oembiometric\ (\calculatePercentage{\oembiometric}{\coem}) are associated with Biometric.
Besides, we failed to categorize 1 case into any group. These OEM feature issues are caused by extra system functions or hardware introduced by manufacturers.

\textbf{Badger}. %
Android 8 (API 26) officially introduces the badge function that shows a dot on a launcher icon. 
However, nearly all OEM-customized launchers support showing a number to indicate the number of notifications instead of only a dot. 
Adaptation methods vary greatly between vendors, requiring different methods to interact with the system.
Xiaomi's badge needs to use Reflection to call an internal method, while Vivo's badge needs to send a Broadcast Intent with specific parameters.

\textbf{Permission}. 
OEMs usually customize permissions and permission management Activities.
When an application requires special permissions, it can use the API \textit{startActivity} to open the settings page and guide the user to grant these permissions~\cite{special-permission}.  
These OEM-defined permissions are usually background limitations, such as the auto-start permission and the battery optimization.
Table~\ref*{tab:rq2-permission} shows the actions of permission management Activity in four OEM systems and the occurrence of these string literals among all repositories in our dataset.

\begin{table}
  \caption{Different Activity action for managing permission}
  \label{tab:rq2-permission}
  \scriptsize
  \setlength{\tabcolsep}{3pt}
  \centering
  \begin{tabular}{lcc}
    \toprule
    Activity Action & OEM & \# of repos \\
    \midrule
    com.huawei.permissionmanager.ui.MainActivity & Huawei & 10 \\
    com.miui.permcenter.permissions.AppPermissionsEditorActivity & Xiaomi & 11 \\
    com.color.safecenter.permission.PermissionManagerActivity & OPPO & 3 \\
    com.sonymobile.cta.SomcCTAMainActivity & Sony & 2 \\
    \bottomrule
\end{tabular}
\end{table}

\textbf{UI}. Manufacturers will also customize the UI.
 In this categorization, we observed that 4 cases are adaptations of the status bar (e.g., RQ1) and 3 cases are adaptations of the notch screen.
Android 9 officially adapts to the notch screen that applications can request to or not to do layout in the notch area~\cite{android-cutout}. 
But Xiaomi has already launched some models equipped with the notch screen, which is based on the Android 8 and uses Xiaomi's adaptation.
However, Xiaomi later abandoned it in new systems based on Android 9 and chose to use the Android 9 official method~\cite{xiaomi-9-cutout}. 
Developers still need to consider Xiaomi's unique adaptation to adapt Xiaomi devices with Android 8 that still exist in the market. %
This example also demonstrates the fragmentation of the Android ecosystem.

\textbf{System Service}. The system services category has issues mainly with vendor-customized and implemented services. 
There are 2 cases are OEM-specific implementations of OAID (Open Anonymous Device Identifier), which is a user-resettable unique identifier like AAID (Android Advertising ID).
Another case from Shouko can be used to customize an extra physical button on the Sony Xperia~\cite{shouko}.

\textbf{Others}. The Camera, Display, and Biometric categories are issues related to additional hardware features.
Developers often use additional parameters to utilize high-performance hardware in these situations.
The biometric library of AndroidX assumes that the biometric module reaches Class 3 security level on some devices. 
Code snippets in the Camera category request higher camera shot parameters on more capable devices.

\textbf{Summary of the OEM Feature category:}
In our analysis of \coem\ OEM issues, we identified several distinct categories, including Badge, Permissions, System Services, UI, Camera, Display, and Biometric issues. 
These issues stem from the introduction of extra system functions or hardware features by OEMs, and their adaptations are often OEM-specific.

\begin{mdframed}[skipabove=3pt]
\textbf{Answer to RQ2:}
Our analysis of Android functionalities impacted by FB issues reveals a broad spectrum of effects, with a notable concentration on Camera and UI issues.
In contrast,
various OEMs typically implement similar additional features beyond the standard Android system, concentrating on additional software features and UI characteristics.
However, different OEMs require app developers to use distinct approaches to implement these features.
\\\textbf{Implications: } Compatibility testing can benefit from analyzing the components affected by real-world compatibility issues.
Although FB issues affect a wide range of functionalities, most of them are still focused on the Camera and UI (73\%). 
Developers can prioritize compatibility testing for these components.
As for OEM features, 
it is useful to create compatibility libraries that abstract the differences in the implementation of similar features across different OEMs.
Such libraries can reduce the developers' workload and improve the consistency of the user experiences across multiple OEMs.
\end{mdframed}

\subsection{RQ3: Issue Fixing Practices}
\subsubsection{Study Method}

In this RQ, we meticulously discussed how developers address FB issues and OEM issues independently by investigating the fixing code.
We reviewed the code for fixing FB issues and found some similar and regular code structures.
We started from a basic observation that FB issues are generally fixed by calling specific APIs.
Therefore, we examined the control/data flow between calling these issue-fixing APIs and checking device information.
More specifically, we investigated how checking device information affects the fixing of compatibility issues in the code structure.
For instance, in Listing~\ref{hw-provider}, the second line checks to see if the current running device is Huawei, and if it is, it tries the workaround if \textit{getUriForFile} fails.
This \textit{if-then-else} pattern is more common in some of the API-Induced compatibility issues~\cite{paper-dhe, paper-xia-rapid}.
However, the code pattern for fixing FB issues is more complex than API-Induced compatibility issues.
In the end, we categorized 5 generic patterns.
We also checked the fix code for the OEM issues.
OEM issues are generally addressed by Reflection or Android Inter-Component Communication (ICC) in a small piece of code.
We examined the specific parameters used in Reflection or ICC.
Besides, we also explored some unusual fixing practices.
In this study, we only considered issues classified as a Functionality break or an OEM feature.
Namely, we did not consider issues that are classified as Other in Table~\ref{tab:rq1-count} of RQ1.
We still applied the same cross-validation approach discussed in Section~\ref{ssec:datacollection} and Section~\ref{sssec:rq1method}.

\subsubsection{Study Results}~\\
Based on our analysis of the dataset, we found that the way to fix DSC issues differs significantly between the functionality break and OEM feature categories.
FB issues are generally solved by calling additional APIs, substituting issue-triggering APIs with alternative APIs, or using a normal approach but with particular parameters.
Addressing OEM feature issues generally leverages Reflection and Android ICC methods to interact with the customized framework.

\textit{Functionality break issues:} Fixing functionality break issues relies heavily on specific patches or workarounds.
These patches usually call or avoid some special API, enable or disable some functionality or change the control flow by overriding a class method.
In general, these patches are trying to get around the misimplemented API.

\begin{figure*}[t!h]
  \centering
  \begin{subfigure}[b]{0.3\linewidth}
    \begin{minted}[fontsize=\footnotesize]{Java}
callAPI1();
if (device) {
    applyWorkaround();
}
callAPI2();
    \end{minted}
    \captionsetup{justification=justified,singlelinecheck=false}
    \caption{Apply workaround}
    \label{fig:fb-pattern-a}
  \end{subfigure}
  \quad
  \begin{subfigure}[b]{0.3\linewidth}
    \begin{minted}[fontsize=\footnotesize]{Java}
if (device) {
    callAlternativeAPI();
} else {
    callAPI();
}
    \end{minted}
    \captionsetup{justification=justified,singlelinecheck=false}
    \caption{Call alternative API}
    \label{fig:fb-pattern-b}
  \end{subfigure}
  \quad
  \begin{subfigure}{0.3\linewidth}
    \begin{minted}[fontsize=\footnotesize]{Java}
bool enableAutoFocus = true;
if (device)
    enableAutoFocus = false;
// ...
callAPI(..., enableAutoFocus, ...); 
    \end{minted}
    \captionsetup{justification=justified,singlelinecheck=false}
    \caption{Specific parameters}
    \label{fig:fb-pattern-c}
  \end{subfigure}
  \bigskip\par
  \begin{subfigure}{0.45\linewidth}
    \begin{minted}[fontsize=\footnotesize]{Java}
bool isFeatureAvailable(feature) {
  if (isProblematicCombination(device, feature)) 
    return false;
  // more combinations...
  return true;
}
    \end{minted}
    \captionsetup{justification=justified,singlelinecheck=false}
    \caption{Disable functionality}
    \label{fig:fb-pattern-d}
  \end{subfigure}
  \quad
  \begin{subfigure}{0.45\linewidth}
    \begin{minted}[fontsize=\footnotesize]{Java}
try {
  callAPI();
} catch(Exception ex) {
  // ignore or try different methods
}
    \end{minted}
    \captionsetup{justification=justified,singlelinecheck=false}
    \caption{Unexpected Exception}
    \label{fig:fb-pattern-e}
  \end{subfigure}
  \bigskip\par
  \caption{General patterns for fixing FB issues}
  \label{fig:fb-class}
\end{figure*}

We categorized the common patterns to fix FB issues into five major categories which we summarized in Figure~\ref{fig:fb-class}.
The \textit{device} if condition in the figures denotes the checking of device information, including both manufacturer and device model.
In addition to using String methods such as \textit{contains}, \textit{equals}, etc. directly in the if statements, developers also utilize helper functions, class fields, and even class polymorphism~\cite{androidx-usetorchasflash} to determine whether the workaround should be applied. This indicates that device information may be propagated across multiple classes, which poses a challenge for designing automated analysis approaches. %

Pattern a describes applying the workaround before calling some API that triggers DSC issues~\cite{androidx-exif-rotation}.
Pattern b indicates that some special approaches are required to replace the original method on particular devices~\cite{androidx-targetaspectratio}, which is very common in fixing API-induced compatibility issues~\cite{paper-dhe}.
Pattern c describes controlling the parameters of the API via device information~\cite{bither-continous-autofocus}.
Pattern d shows the utility function used to mask special functions on multiple devices~\cite{androidx-usetorchasflash-impl}.
Patterns c and d are usually used to disable problematic features, as the example we show in Listing~\ref{lst-exoplayer}.
Although pattern d can eventually lead to calling different APIs (pattern a and b) or using different parameters (pattern c), we distinguished it from others because it usually checks multiple combinations of devices and features and is commonly used in large-scale compatibility libraries, such as CameraX and ExoPlayer~\cite{exoplayer}.
Pattern e is a special pattern that handles unexpected Exceptions when calling specific APIs. This is because a few buggy APIs will throw unexpected Exceptions, which will crash the application if not captured by the try-catch. Examples of each pattern are available in our artifacts.

We counted the usage of the five patterns for fixing FB issues in Table~\ref{tab:rq3-fb-pattern}.
Most FB issues (38\%) are fixed by pattern a, and pattern b accounts for 15\%.
Using different API parameters (pattern c) accounts for 24\%.
Creating a dedicated helper function to check the feature's availability (pattern d) accounts for 20\%, which is relatively common in large-scale repositories.
CameraX contributes to 18 cases (62\%) among this pattern.
Besides, very few FB issues (2\%) require additional try-catch statements (pattern e).
Camera issues are mostly solved by patterns a, c, and d (87\%).
This is because only some of the camera features need to be masked.
UI issues are usually addressed by patterns a and b (74\%).

\begin{table}
  \caption{Occurrences for each pattern on fixing FB issues}
  \label{tab:rq3-fb-pattern}
  \setlength{\tabcolsep}{4pt}
  \footnotesize
  \centering
  \begin{tabular}{cccccc}
    \toprule
    Category         & Pattern a & Pattern b & Pattern c & Pattern d & Pattern e \\
    \midrule
    Camera           & 29        & 9         & 23        & 20        & 2         \\
    UI               & 9         & 8         & 4         & 2         & 0         \\
    Others FB issues & 16        & 5         & 7         & 7         & 1         \\
    Total            & 54        & 22        & 34        & 29        & 3         \\
    \bottomrule
  \end{tabular}
\end{table}

\textit{OEM feature issues:}  %
 In most cases, \textbf{adapting OEM features cannot be achieved by calling the OEM-customized functions directly} because the application is compiled using the standard Android SDK, which does not have the OEM-customized parts.
For UI-related features, it may be more common to leverage Reflection to access internal methods or fields because the applications need to modify specific properties of the UI object.
Examples include the adaptation methods of the status bar (Listing~\ref{lst-miui}) and the notch screen.
62\% (8 of 13) Badge OEM issues and 66\% (6 of 9) UI OEM issues are addressed through Reflection.

OEM features implemented with Activity, Broadcast or ContentResolver need to be accessed using Intent or Uri with a particular action or path string.
\textbf{These string literals typically contain distinctive OEM information, such as brand or system name.}
For example, the Intent actions of the Activity used to manage permissions in different systems in Table~\ref{tab:rq2-permission} contain their brand names.
Among the Badger OEM issues, Broadcast is used in 8 issues, while 6 issues from Xiaomi's adaptation methods try both Reflection and Broadcast.
In the Permission OEM issues, 58\% (7 of 12) are implemented with \textit{startActivity} to open the permission management Activity. While others utilize Reflection and deprecated Android API to check whether specific permission is granted.
 Also, the names of classes, fields, and methods used in Reflection usually contain OEM information.

Besides, \textbf{the parameters of the adaptation methods vary from vendor to vendor}, which makes it difficult for developers to reuse existing implementations to adapt to multiple vendors.
Thus, developers usually create independent methods for different vendors and choose to call them by checking the running device information.
Listing~\ref{lst-api-comp} is an example of using the abstract class and inherit to adapt to multiple devices.

Despite the above adaptation methods, we also found a way to adapt OEM features by configuring key-value properties and declaring OEM-defined permissions in the \textit{AndroidManifest.xml} file.
We found one case related to Samsung's Dex Mode in our dataset.
The Dex Mode is a multi-monitor multitasking windowed display mechanism launched by Samsung~\cite{samsung-dex-mode}.
Samsung requires applications to enable \textit{resizeableActivity}, which is an attribute defined by Android, and declare meta-data with a special key to handle runtime configuration changes as an enhancement.
We searched for the key prefix \textit{com.samsung.android} in all AndroidManifest files, and we found 8 applications in our Stage I repositories set that have adapted to this feature.
Besides, Xiaomi offers a pedometer service~\cite{xiaomi-doc-steps}.
Applications need to declare a Xiaomi-specified permission in the \textit{AndroidManifest} and use the ContentProvider to access system-provided data.
This may enable a customized system of more fine-grained permission management. %

Determining control flow by device-related properties and executing device-specific codes is still the primary way to resolve DSC issues.
However, not all fixes involve checking device information.
There is an application that avoids Huawei's destruction of the API contract by rewriting the \textit{getExternalFilesDir} method but does not use the \textit{Build} class to check whether it is running on Huawei devices~\cite{hw-repo-3}.
We also observed similar behavior in the OEM features category.
There is another repository that maintains a list of all possible paths to the permission management Activity used by different OEMs and uses \textit{resolveActivityInfo} to try each one without checking the running device~\cite{stopcovi59:online}.
This may bring challenges to subsequent automated analyses of DSC issues.

\begin{mdframed}[skipabove=3pt]
  \textbf{Answer to RQ3:}
  Our analysis reveals notable differences in addressing functionality break and OEM feature issues.
  For FB issues, common solutions include calling additional APIs, substituting problematic APIs with alternatives, or using standard approaches with specific parameters.
  Resolving OEM feature issues frequently leverages Android inter-component communication methods and Reflection.
  We also revealed some unconventional strategies that adapt OEM features and permissions in the AndroidManifest and address compatibility issues without checking the running device model.
  \\\textbf{Implications:} Our study on repair methods can better support the automated extraction, analysis, and repair of device-specific compatibility issues in the future.
  Our analysis of the fixing patterns of FB issues also suggests that they are more complex compared to evolution-induced API compatibility issues.
  We revealed that the adaptation of OEM features often uses string constants containing OEM information, which can be a key point in extracting OEM issue fixings. %

\end{mdframed}

\section{Threats to Validity}

To reduce human error, we applied the widely used cross-validation to ensure that each classification result had been checked by at least two authors and agreed upon.
Our manual checks also implemented cross-validation through code commit records, and additional information from StackOverflow and Google Issue Tracker, which helped us improve the accuracy of our identification.

Our dataset was collected by searching for references to the \texttt{Build} class.
But applications might also address device-specific issues without accessing the class, causing our dataset to be unable to collect such code snippets.
For example, applications may utilize \texttt{SystemProperties} to retrieve other device-related information that is not included in the \texttt{Build} class~\cite{stackof-android-properties}. But this approach is not recommended by Android.
Besides, We have already shown in RQ3 that it is possible to handle device-specific issues without checking the running device.
Our search criteria cannot cover these situations.
Moreover, our approach cannot capture cases where applications utilize external compatibility libraries to address DSC issues. However, these libraries still access device information through the Build class, and our dataset also includes several compatibility libraries, such as AndroidX.

In this paper, we only analyzed open-source Android apps.
Proprietary and closed-source apps may contain additional fixes for DSC issues.
Our study may not capture the characteristics of these issues.
We limited the dataset to open-source repositories because they contain substantial information for us to identify and understand these issues.
To mitigate the problem, we collected repositories that are popular and well-maintained, including libraries maintained by Google that are commonly used in closed-source Android apps.

\begin{table}
    \caption{Existing Work about Android Compatibility Issue}
    \label{tab:existing-work}
    \centering
    \begin{tabular}{lcc}
        \toprule
        Existing Work & Proposed Tool & \multicolumn{1}{p{7em}}{\centering Related to \\ DSC Issue} \\
        \midrule
        WCRE'2012, Han et al.~\cite{paper-han2012} & - & Yes \\
        ASE'2016, Wei et al.~\cite{paper-taming} & FicFinder & Yes \\
        ASE'2018, He et al.~\cite{paper-dhe} & IctApiFinder & \\
        ISSTA'2018, Li et al.~\cite{paper-cid} & Cid & \\
        ICSE'2019, Wei et al.~\cite{paper-pivot} & Pivot & Yes \\
        ICSE'2020, Xia et al.~\cite{paper-xia-rapid} & RAPID & \\
        ICSE'2022, Zhao et al.~\cite{paper-zhao2022-reparing} & RepairDroid & Yes \\
        ISSTA'2022, Liu et al.~\cite{paper-liu-replicability} & - & Yes \\
        TOSEM'2023, Liu et al.~\cite{paper-liu-tosem23-auto} & AndroMevol & Yes \\        
        \bottomrule
    \end{tabular}
\end{table}

\section{Related Work}

\textbf{Detecting and Fixing Compatibility Issue}. 
Various aspects of Android compatibility issues were studied in existing research.
Table~\ref{tab:existing-work} lists existing work regarding the Android compatibility issues.
Han et al. first highlighted Android's fragmentation from the perspective of vendor-specific bug reports from HTC and Motorola~\cite{paper-han2012}.
Wel et al. provided an in-depth analysis of fragmentation-induced compatibility issues and presented both device-specific and non-device-specific (API-related) compatibility issues~\cite{paper-taming,paper-tse-2020}.
They proposed a prototype tool \textsc{FicFinder} to detect such issues.
This is the work that is the closest to ours.
Wei et al.'s work has several limitations. First, it analyzed 110 device-specific compatibility issues in only five apps. This challenges the generalizability of their study. 
Second, their key findings are drawn based on both device-specific and non-device-specific issues.
This prevents Wei et al. from conducting a thorough investigation into features of device-specific issues.
Their key conclusions are drawn based on general compatibility issues and cannot hold for DSC issues. For example, they concluded that patches for compatibility issues are usually simple. Meanwhile, our study exposed that patches for DSC issues are complex and vary across issues (RQ3). Their general findings cannot support the effective detection of DSC issues.
In contrast, we conduct a concentrated study that extensively characterizes device-specific compatibility issues.
We make distinct observations that device-specific issues can be further divided into functionalities breaks and OEM features and the nature of these two types is different.
We also observe that breaks in a few Android functionalities (i.e., Camera and UI) account for the majority of the compatibility issues.
Such in-depth insights cannot be made in Wei et al.'s study on general Android compatibility issues.
Besides, the \textsc{FicFinder} they designed relied on the pre-defined patterns (API-Context pairs) to detect possible occurrences of compatibility issues.
Later, Wei et al. launched \textsc{Pivot}, aiming to \replace{extract}{learn} API-device correlations of device-specific compatibility issues from large Android app corpus~\cite{paper-pivot}.
However, this work did not provide additional insights on device-specific compatibility issues.

Other studies delve into the compatibility issues caused by the evolution of Android APIs.
He et al. developed the \textit{IctApiFinder}~\cite{paper-dhe} to detect incompatible API usages in Android applications.
Li et al. introduced \textsc{CiD}~\cite{paper-cid} as an automated method to model API lifecycles and identify potential issues related to APIs in app code.
Both of their work relied on API availability across multiple Android versions, which can be easily retrieved through Android documentation and comparisons on Android framework code of different versions.
However, this situation does not exist in our study.
Furthermore, their approaches check whether an API may be called on an incompatible runtime, which will throw a \textit{NoSuchMethod/FieldError}.
Liu et al. proposed \textit{AndroMevol}~\cite{paper-liu-tosem23-auto} to harvest incompatible methods and fields in multiple vendors' systems, but they were limited to the non-existent fields/methods and did not address the characteristics of DSC issues.
In addition to learning about incompatible APIs from the Android framework, the RAPID~\cite{paper-xia-rapid} approach proposed by Xia et al. learned such information from applications.
Zhao et al. put forward \textsc{RepairDroid} to create fix templates for compatibility issues towards automatically repairing~\cite{paper-zhao2022-reparing}.
Other studies have also perceived the incompatibility of API implementations or API Semantics.
Huang et al. focused on compatibility issues of callback APIs due to Android evolution~\cite{paper-callback}.
Huang et al. also noticed the compatibility issues in XML configuration files and proposed \textsc{ConfDroid} to detect and \textsc{ConfFix} to repair them~\cite{paper-confdroid, paper-conffix}.

Our study scope is different from these papers.
We focus on device-specific compatibility issues.
Due to their different causes, device-specific compatibility issues and API-evolution-induced compatibility issues are different by nature.
For example, API-evolution-induced compatibility issues are often documented and they are fixed in a unified way.
Nevertheless, device-specific compatibility issues are usually not documented and the fixes can be divergent.

\textbf{Compatibility Testing}.
Researchers have also tried to mitigate device-specific compatibility issues using testing approaches.
Vilkomir et al. and Khalid et al. prioritized representative devices to test as a way to reduce the cost of large-scale testing~\cite{vilkomir-testing-selection,khalid2014-Prioritizing-testing}.
Fazzini et al. proposed \textsc{DiffDroid} to detect inconsistent UI displays on different Android devices~\cite{paper-cpi}.
Ki et al. designed \textsc{Mimic} specifically for differentiating UI behaviors across different devices~\cite{paper-ki-mimic}.
The findings in our paper can better guide the design of testing techniques to expose device-specific compatibility issues.

\section{Conclusion}
This paper presents an empirical study on device-specific compatibility issues in Android apps.
We analyzed \fictotal\ compatibility issues in \repototal\ from a new perspective by categorizing them into functionality breaks and OEM features.
Our study shows that although a variety of Android functionalities can be broken due to device customization, the majority (68.5\%) of the issues stem from Camera or UI functionalities.
This observation implies that compatibility testing efforts should be prioritized for app modules involving these functionalities.
We also analyzed the fixing practices of device-specific compatibility issues and disclosed the distinct approaches to resolve functionality break issues and OEM feature issues.
Particularly, in dealing with OEM features, the use of Reflection and inter-component communication is common, and their parameters often exhibit significant OEM-specific characteristics.
To summarize, we conducted an in-depth study characterizing device-specific compatibility issues in the source code of Android apps.
Our results characterize the root causes and manifestation of compatibility issues and shed light on future directions to improve compatibility issue detecting and fixing practices.

\section{Data Availability}
The list of analyzed repositories, collected issues, the scripts to collect the dataset, the backup of analyzed repositories, and examples for RQ.3 are available at:
\url{https://zenodo.org/doi/10.5281/zenodo.10531780}

\section{Acknowledgments}

The authors thank ICSME 2024 reviewers for their constructive comments.
This work is partially supported by the National Natural Science Foundation of China (Grant Nos. 61932021 and 62372219), and the NSERC Discovery Grant RGPIN-2022-03744 DGECR-2022-00378.

\IEEEtriggeratref{61}
\bibliographystyle{IEEEtran}
\bibliography{references}

\end{document}